\documentclass[aps,prl,reprint,twocolumn,showpacs,floatfix,superscriptaddress]{revtex4-1}

\usepackage{amssymb,amsmath,amstext}                
\usepackage{graphicx}                                               
\usepackage{epstopdf}                                               
\usepackage{color}                                                     
\usepackage{bm}                                                        
\usepackage{appendix}                                              
\usepackage[utf8]{inputenc}
\usepackage{latexsym}
\usepackage{xcolor}
\usepackage{braket}
\usepackage{ulem}
\definecolor{lblue} {RGB}{51,71,158}
\usepackage[colorlinks=true,citecolor=blue,linkcolor=blue,urlcolor=lblue]{hyperref}

\begin{document}

\title{Thouless time analysis of Anderson and many-body localization transitions}
\author{Piotr Sierant}
\affiliation{Instytut Fizyki im. Mariana Smoluchowskiego, Uniwersytet Jagiello\'nski,  \L{}ojasiewicza 11, 30-348 Krak\'ow, Poland }
\email{piotr.sierant@uj.edu.pl}

\author{Dominique Delande}
\affiliation{Laboratoire Kastler Brossel, Sorbonne Universit\'e, CNRS,
ENS-PSL Research University, Coll\`ege de France, 4 Place Jussieu, 75005
Paris, France}

\author{Jakub Zakrzewski}
\affiliation{Instytut Fizyki im. Mariana Smoluchowskiego, Uniwersytet Jagiello\'nski,  \L{}ojasiewicza 11, 30-348 Krak\'ow, Poland }
\affiliation{Mark Kac Complex
Systems Research Center, Uniwersytet Jagiello\'nski, Krak\'ow,
Poland. }

\date{\today}


\begin{abstract}
Spectral statistics of disordered systems encode   
   Thouless and Heisenberg time scales whose
ratio determines whether the system is chaotic or localized.
We show that the scaling of the Thouless time with system size and disorder strength is very similar in one-body Anderson models and in disordered  quantum many-body systems. 
We argue that the two-parameter scaling breaks down 
in the vicinity of the transition to the localized phase signaling slow down of dynamics. 
\end{abstract}

\maketitle

{\it Introduction. } 
  The phenomenon of many-body localization (MBL)  \cite{Gornyi05, Basko06}, the robust mechanism of ergodicity 
     breaking in quantum world \cite{Nandkishore15, Alet18, Abanin19} has received a lot of 
     attention over the last decade.
     Investigations  of MBL in lattice models, pioneered in spin systems \cite{Santos04a, Oganesyan07,Pal10},
     were extended to bosonic models \cite{Sierant18, Orell19}
     and to systems of spinful fermions \cite{Mondaini15, Prelovsek16, Zakrzewski18, Kozarzewski18}.
     Remarkably, MBL, usually thought of as Anderson 
     localization \cite{Anderson58} in presence of interactions, was shown to occur in systems with completely delocalized single particle
     states either due to random interactions \cite{Sierant17, Lev16, Li17a}
     or in a quasiperiodic Fibonacci chain \cite{Mace19}.
     MBL was also found in disorder-free systems as a result of gauge invariance
     \cite{Smith17, Brenes18} or due to Wannier-Stark localization \cite{Schulz19,vanNieuwenburg19}, 
     in systems with power-law interactions
     \cite{DeTomasi18,Safavi19,Botzung18,Maksymov19a}, or even with an infinite range \cite{Sierant19c} as well as
    driven Floquet MBL systems \cite{Bordia17}.
     Local integrals of motion
     \cite{Serbyn13b,Huse14,Ros15,Imbrie16, Wahl17, Mierzejewski18, Thomson18}
     provide a common framework to understand features of MBL such as area-law entanglement entropy of eigenstates \cite{Bauer13,Serbyn13a}, 
     logarithmic growth
     of bipartite entanglement entropy after quench from a separable state \cite{Znidaric08, Bardarson12} or
     Poisson statistics (PS) of energy levels.

   The crossover between level statistics of an ergodic system with time reversal symmetry which follow predictions of the
    Gaussian Orthogonal Ensemble (GOE) of random matrices  \cite{Mehtabook,Haake} and PS of MBL phase seems 
   to be well understood \cite{Serbyn16,Bertrand16,Kjall18, Buijsman18, Sierant19b, Sierant19d}.
   However, a recent analysis \cite{Suntais19} of the Spectral Form Factor (SFF), $K(\tau)$,
    in the wide regime of slow thermalization  on the ergodic side of the crossover
    \cite{BarLev15, Luitz16, Luitz16b, Mierzejewski16} questions the very existence of the MBL phase in 
    the thermodynamic limit predicting a two-parameter scaling of Thouless time
    \begin{equation}
     t_{Th} =t_0 \mathrm{e}^{W/\Omega }L^2,
     \label{ThSc}
    \end{equation}
    where $L$ is system size, $W$ is disorder strength, $t_0$ and $\Omega$ are constants. 
    The Thouless time $t_{Th}$ 
    is defined as the time scale beyond
    which the SFF follows the universal GOE form.
    Another important time scale, the Heisenberg time $t_H=2\pi/\Delta$ is defined by the average 
    level spacing $\Delta$ which scales exponentially with a many-body system size $L$, $\Delta \propto e^{cL}$.
    The Heisenberg time $t_H$ is  a limit
    beyond which the discrete nature of the energy spectrum manifests itself and where 
    system dependent quantum effects are unavoidable. In the thermodynamic limit, \eqref{ThSc}
    implies $t_{Th}/t_H\rightarrow0$.
    Hence, \cite{Suntais19} arrives at the surprising conclusion that disordered quantum spin chains 
have spectral properties following the GOE 
predictions
regardless of the disorder strength $W$ and that 
MBL 
is merely a finite-size effect.

In this letter 
we analyse the SFF in the delocalized phase and its modifications when approaching the transition  to the localized phase. We show that the Thouless time scales like $L^2$, in agreement with \eqref{ThSc}, in the deep delocalized phase in Anderson models as well as in disordered many-body systems. The scaling with $L$ evolves to a larger power at the critical point of Anderson model, a phenomenon that we correlate with the  diffusive and subdiffusive transport properties respectively in the delocalized phase and at the metal-insulator transition. 
Results obtained for 3D and 5D Anderson models with known localization properties put the conclusions of \cite{Suntais19} about the scaling of Thouless time $t_{Th}$ 
 in a considerable doubt, suggesting the presence of a MBL phase at sufficiently strong disorder when finite-size effects are properly taken into account.

{\it Thouless time.}
  In a non-interacting system, the Thouless time  was introduced as 
the time to diffuse through the system and reach its boundary \cite{Thouless74}.
It determines the energy scale below which the level statistics
are well described by GOE \cite{Shklovskii93}, 
whereas its ratio with the Heisenberg time fixes the dimensionless 
conductance of the system \cite{Edwards72} and enters the scaling theory of Anderson localization 
transition \cite{Abrahams79}.  
The Thouless time $t_{Th}$ in disordered many-body systems
can be probed by examining the behavior of the SFF
\cite{Cotler17, Chen18, Gharibyan18, Chan18} defined as
     \begin{equation}
 K(\tau)  = \frac{1}{Z}\left \langle \left| \sum_{j=1}^{\mathcal N} g( \epsilon_j) \mathrm{e}^{-i \epsilon_j \tau} \right|^2 \right \rangle,
 \label{eq: Kt}
\end{equation}
where $\epsilon_j$ are eigenvalues of the system after the unfolding \cite{Gomez02} (which sets 
their density to unity), 
$g(\epsilon)$ is a Gaussian function 
reducing influence  of
the spectrum's edges, the average is taken over
disorder realizations and $\mathcal N$ is the dimension of the Hilbert space.
For a GOE matrix, the SFF is known analytically: $K_{GOE}(\tau)  = 2 \tau - \tau \log(1+2\tau)$
for $\tau \leqslant 1$ and $K_{GOE}(\tau)  = 2 \tau - \tau \log(1+2\tau)$ for $\tau > 1$.
The linear ramp $K_{GOE}(\tau)\approx2\tau$ of SFF starting at $\tau=0$ 
reflects correlations between all pairs of eigenvalues in a GOE matrix. 
In contrast, SFF $K(\tau)$ calculated for a physical system follows the GOE predictions
$K(\tau)=K_{GOE}(\tau)$ only for $\tau > \tau_{Th}$ defining $\tau_{Th}$, 
which, in turn, is proportional to the Thouless time
$t_{Th} = \tau_{Th} t_H$. The proportionality factor $t_H$ comes from the fact that 
unfolded eigenvalues $\epsilon_i$ enter the definition of $K(\tau)$; it is equal to the Heisenberg time $t_H$, 
determined by the inverse level spacing.

For a diffusive transport, the mean square displacement $\langle r^2(t) \rangle$ is proportional to time $t$. Hence, the above definition of  $t_{Th}$ coincides with 
the original definition of Thouless time in diffusive system provided that the $t_{Th}\sim L^2$ where $L$ is the system size. For subdiffusion, the mean square displacement behaves as $\langle r^2(t) \rangle \sim t^{\alpha}$ with $0 <\alpha < 1$, thus we expect $t_{Th}\sim L^{2/\alpha}$.
 In the deeply localized regime where the localization length is much smaller than the system size, a particle never explores the full system size, so that the original Thouless time eventually diverges and becomes larger than the Heisenberg time. In contrast, Poisson statistics are characteristic for
the localized regime where the SFF is independent of time; the Thouless time deduced from the SFF is thus equal to the Heisenberg time. This implies that the latter definition 
is applicable only in the delocalized regime. 
 Before we consider interacting models,
we examine first the Thouless time as defined by the SFF 
in Anderson models.

{\it Thouless time in 3D and 5D Anderson models.} 
    The Hamiltonian of the Anderson model describes hopping of a particle on a $D$-dimensional 
lattice with disorder and reads
\begin{equation}
 \hat H = -t\sum_{\langle i,j\rangle} (\hat c^{\dag}_i \hat c_j + h.c.) + \sum_i \epsilon_i \hat c^{\dag}_i \hat c_i,
 \label{eqand1}
\end{equation}
   where $\hat c^{\dag}_i$ is creation operator for particle at site $i$,
$\langle .,.\rangle$ denotes sum over neighboring lattice sites,
$t\equiv1$ is the tunneling amplitude and $\epsilon_i\in [-W,W]$ 
are uniformly distributed uncorrelated random variables forming
on-site potential. 
  \begin{figure}
 \includegraphics[width=0.92\linewidth]{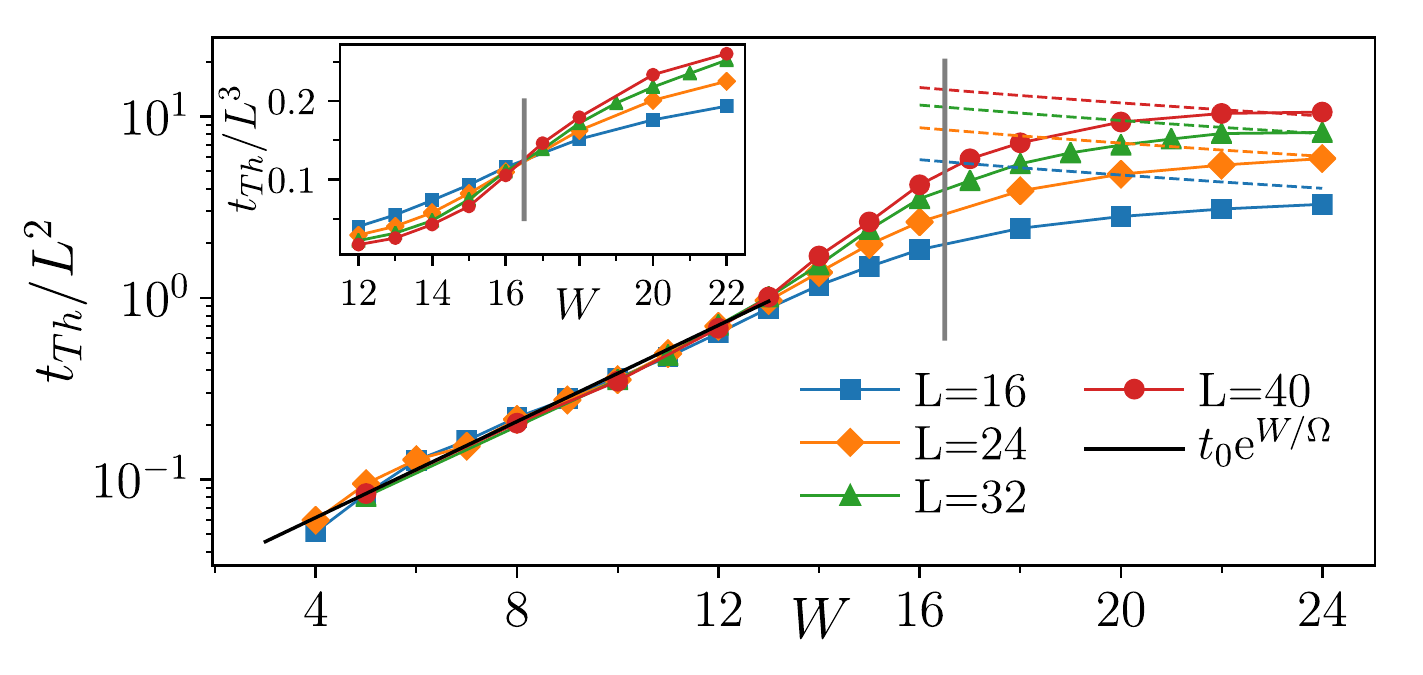} \\ \vspace{-0.15cm}
  \includegraphics[width=0.92\linewidth]{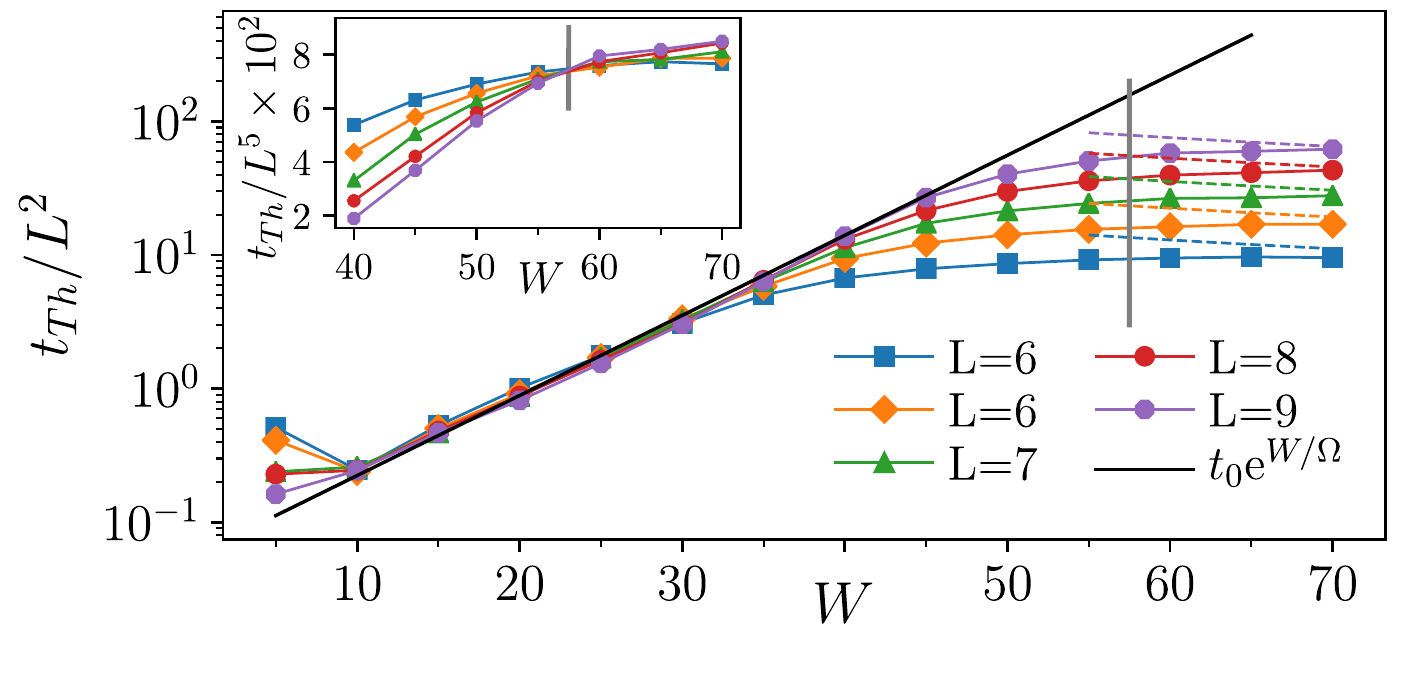} \vspace{-0.65cm}
  \caption{  Thouless time $t_{Th}$
  vs. disorder strength $W$ extracted from the SFF for 3D (upper plot) and 5D (lower plot) Anderson models, for various system sizes $L$. The black solid lines denote the scaling of Eq.~\eqref{ThSc}, grey vertical lines denote the critical disorder strength $W^{3D}_C=16.54$ ($W^{5D}_C=57.3$)
  in 3D (5D) model. Dashed lines denote the Heisenberg time $t_H$. The insets show $t_{Th}/L^3$ ($t_{Th}/L^5$) in the 3D (5D) case. 
  }\label{andtTh}
\end{figure}     
Numerical studies of transport properties of the 3D Anderson model
\cite{MacKinnon81, MacKinnon83, Kramer90, Kramer93} indicate 
that transport is diffusive for disorder strengths
$W<W^{3D}_C\approx 16.54$ \cite{Slevin18} and that the system
remains insulating for $W>W_C$. 
Exactly at the transition, the 3D Anderson model is characterized by subdiffusion \cite{Ohtsuki97} and multifractal wave functions \cite{Rodriguez09, Rodriguez10}.
  Studies of transport in 5D Anderson model \cite{Ueoka14} 
find a localization transition, 
consistently with studies of level statistics \cite{Garcia07} giving the critical disorder
$W^{5D}_C = 57.3,$ confirmed in~\cite{Tarquini17, Pietracaprina16}.

  Level spacing distribution in the 3D Anderson model were studied in
\cite{Shklovskii93, Hofstetter93, Zharekeshev95, Varga95, Zharekeshev97}. 
Thouless times presented in Fig.~\ref{andtTh} unveil
a long-range correlation aspect of level statistics in Anderson models.
Examples of SFF and details on Thouless time estimation are 
given in \cite{suppl}.

At small disorder strength $W$, the Thouless times 
depend quadratically on system size $L$ (Fig.~\ref{andtTh}), following precisely the scaling \eqref{ThSc} 
which simply means that the dynamics is diffusive.
For the 3D model, the $t_{Th}/L^2 \propto e^{W/\Omega}$ behavior persists
up to $W\approx 12$. For bigger disorder strength, the quadratic scaling with 
the system size is no longer valid. Directly at the transition, $W=W^{3D}_C$,
the Thouless time should scale as the Heisenberg time i.e. $t_{Th} \propto L^3$.
This is indeed the case as the inset in the upper plot in Fig.~\ref{andtTh}
demonstrates. Further increase of the disorder strength leads to a slow increase 
of the Thouless time $t_{Th}$ with eventual saturation to the 
Heisenberg time $t_H$. 

In the deep delocalized phase where $t_{Th}$ scales with $L^2$, 
the ratio  $t_{Th}/L^2$ is nothing but - up to a constant multiplicative
factor - the inverse of the diffusion coefficient $D(W),$ in accordance
with the original definition of the Thouless time. The dependence of 
$D(W)$ with $W$ is not known analytically, but it is known that it 
decreases quickly with $W,$ vanishing at the critical point and 
scaling like $(W_c-W)^s$ below it, with the critical exponent 
$s\approx 1.574.$ In any case, it is definitely not $e^{-W/\Omega}$ 
as  in \eqref{ThSc}. It may be that, in a limited
range of $W$ values, $D(W)$ can be approximately fitted by
an exponential decrease, but other forms could do the job as well.

The 5D case is essentially identical, except that the
Thouless time scales like $L^5$ instead of $L^3$ at the critical point.
The growth of the Hilbert space size as $L^5$ prevents reaching system sizes $L\geq10$. Nevertheless, the obtained 
Thouless times $t_{Th}$, when rescaled by $L^5$ as suggested 
by the relation $t_{Th} \sim t_H$ valid at the transition, 
lead to a clear crossing of the $t_{Th}/L^5$ curves at $W^{5D}_C$.
The $W$ dependence of $t_{Th}/L^2$ in the deeply delocalized 
regime is again approximately reproduced by an exponential,
although it certainly fails near the critical point.

\begin{figure}
 \includegraphics[width=1.0\linewidth]{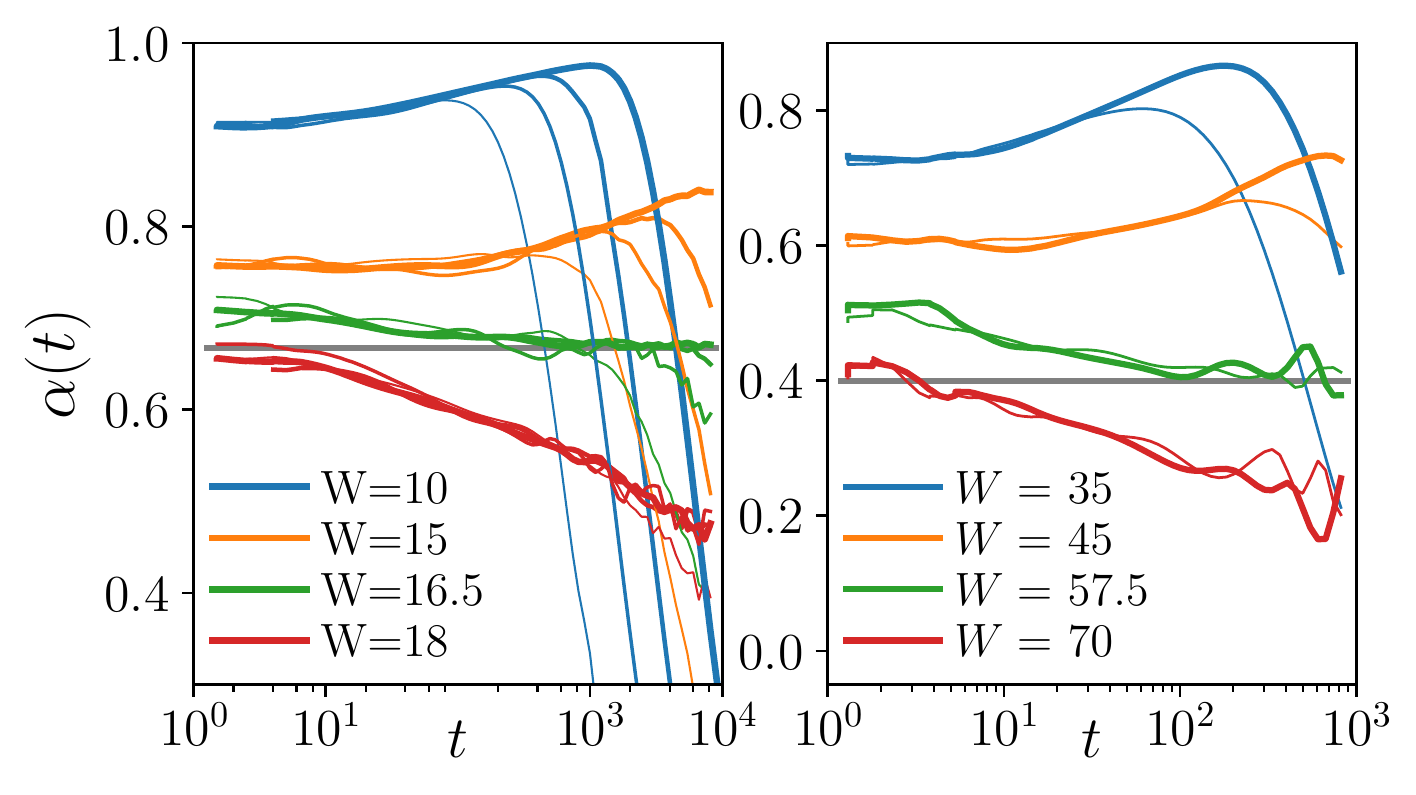} \vspace{-0.79cm}
  \caption{ { Time dependent $\alpha(t)$ function
  for 3D (left) and 5D (right) Anderson models for various
  disorder strengths $W$. In the 3D case, results for
  the system size {$L=80,120,160,240$} 
  are denoted by {progressively thicker} lines,
  whereas
  in the 5D case {thin (thick)} lines correspond to $L=20$ ($L=30$). 
 }} \label{andB} 
\end{figure} 

 {\it Diffusion and subdiffusion in Anderson models.} 
To demonstrate that the obtained behaviors of the Thouless
time $t_{Th}$ are related to time dynamics in Anderson systems, we consider the initial state 
$|\psi_0\rangle$ with a  particle located 
at a given lattice site with periodic boundary conditions. 
The time evolved state $|\psi_0(t) \rangle = e^{-i \hat H t} |\psi_0\rangle$ 
is obtained  employing the Chebyshev technique \cite{Fehske08} 
that allows us to get results for system sizes up to $L=240$ and $L=30$ 
for the 3D and 5D cases, respectively. The mean square displacement
\begin{equation}
 \langle r^2(t)\rangle = \langle \psi_0(t) | \sum_{i=1}^{D}  (\hat r_i - \overline r_i )^2|\psi_0(t)\rangle, 
\end{equation}
where $r_i$ is $i$-th component of the position operator $\bf{ \hat r}$ and $\overline r_i = \langle \psi_0(t) | \hat r_i |\psi_0(t)\rangle$, allows us   to distinguish 
(considering first the $L\rightarrow \infty$ limit and then looking at times $t \gg 1$):
diffusive $ \langle r^2(t)\rangle \propto D t$, subdiffusive  $ \langle r^2(t)\rangle \propto t^{\alpha}$ and localized behaviors. The latter occurs when $ \langle r^2(t)\rangle$ saturates after the initial expansion of the wave packet.
Time dependence of the mean square displacement is  reflected by the function
$\alpha(t) \equiv d \log \langle r^2(t) \rangle / d \log t$.
In the case of diffusion $\alpha(t)=1$, for subdiffusion $0<\alpha(t)=\alpha<1$
and in the localized case $\alpha(t) \rightarrow 0$.

On the delocalized side of the transition in 3D and 5D  models, respectively for $W<W^{3D}_C$ and $W<W^{5D}_C$, we observe (Fig.~\ref{andB}) that $\alpha(t)$ initially increases over time
reaching larger maximal values for increasing system sizes. Assuming that this trend persists with increasing system size, taking the thermodynamic limit $L\rightarrow\infty$ we end up with diffusive behavior $\alpha(t)=1$ for $t \gg 1$. 
The decrease of $\alpha(t)$ observed at the delocalized side of the transition for a given system size $L$ occurs when the wave packet ceases to spread as its size approaches the system size. The situation is different at the transition, where, regardless of the system size, $\alpha(t)$ approaches a constant value $\alpha_{3D} = 2/3$ in the 3D case \cite{Ohtsuki97, Lemarie10} 
or $\alpha_{5D}=2/5$ in the 5D case. Subsequently, $\alpha(t)$ decreases when the size of wavepacket approaches the system size $L$. This indicates that in the thermodynamic limit $L\rightarrow \infty$, for $t\gg1$, there is a subdiffusion $\alpha(t)\rightarrow \alpha_{3D} (\alpha_{5D})$ at the transition in 3D (5D) Anderson model. Finally, for $W>W^{3D}_C$ ($W^{5D}_C$), $\alpha(t)$ decreases with time being nearly independent of the system size -- a sign of localization. 

The observed diffusion and subdiffusion for both 3D and 5D  models agree with results obtained for the Thouless time $t_{Th}$. In diffusive system $\langle r(t)^2 \rangle \propto D t$ which means that the time for reaching the boundary of the system $t^B_{Th} \propto L^2$. For subdiffusion, $\langle r(t)^2 \rangle \propto t^{\alpha}$ implies that  $t^B_{Th} \propto L^{2/\alpha}$. Given the values for $\alpha_{3D}$ and $\alpha_{5D}$ we see that the obtained scalings of $t^B_{Th}$ on the delocalized side of the transition and at the transition agree with the scalings $t_{Th}\propto L^2$ and $t_{Th}\propto L^3$ (or $t_{Th}\propto L^5$ in the 5D case) obtained from the SFF. 
  
  The results shown in Fig.~\ref{andB} highlight the importance of finite size and finite time effects. The limit $L \rightarrow \infty$ followed by
   $t\rightarrow \infty$ has to be carefully examined to reveal the trend towards diffusion/subdiffusion in the system. 
    For instance, if data for 3D  model at $W=15$ were available only up to time $t=10^2$ one could incorrectly assume a subdiffusion with $\alpha \approx 0.75$. It seems plausible that the case of interacting systems is analogous suggesting that the claims about subdiffusion on the ergodic side of MBL transition \cite{BarLev15, Agarwal15, Luitz16, Varma17} might be invalid in the asymptotic limit $L\rightarrow \infty$, $t\gg1$  \cite{Bera17, Weiner19}.

 {\it Thouless time in disordered many-body systems.}  
    Consider 1D disordered spin-1/2 chains with Hamiltonian:
\begin{eqnarray}
 \nonumber 
 H= J_1 \sum_{i=1}^{L} \left( S^x_{i}S^x_{i+1}+S^y_{i}S^y_{i+1} + \Delta S^z_{i}S^z_{i+1}  \right) + \sum_{i=1}^{L} h_i S^z_i
 \\  
 +J_2 \sum_{i=1}^{L} \left( S^x_{i}S^x_{i+2}+S^y_{i}S^y_{i+2} + \Delta S^z_{i}S^z_{i+2}  \right), \quad
 \label{eq: XXZ}
\end{eqnarray}
where  $\vec{S}_i$ are spin-1/2 matrices, $J_1=1$ is  the energy unit,
periodic boundary conditions are assumed and $h_i \in [-W, W]$ 
are independent, uniformly distributed random variables.
Setting $J_2 = 0$ and $\Delta = 1$  we arrive at a disordered XXZ model,
widely studied in the MBL context 
\cite{Berkelbach10,  Bera15, Enss17, Herviou19, Colmenarez19, Chanda19, Pietracaprina18}, in particular, an analysis of mean gap ratio $\overline r$  \cite{Atas13} predicts the critical value of disorder strength $W_C = 3.72(6)$ \cite{Luitz15}
for transition to a MBL phase. Similar reasoning leads to $W_C\approx 9$ for $J_1-J_2$ model.
For details of our calculations of Thouless times see \cite{suppl}.

 \begin{figure}
 \includegraphics[width=0.50\linewidth]{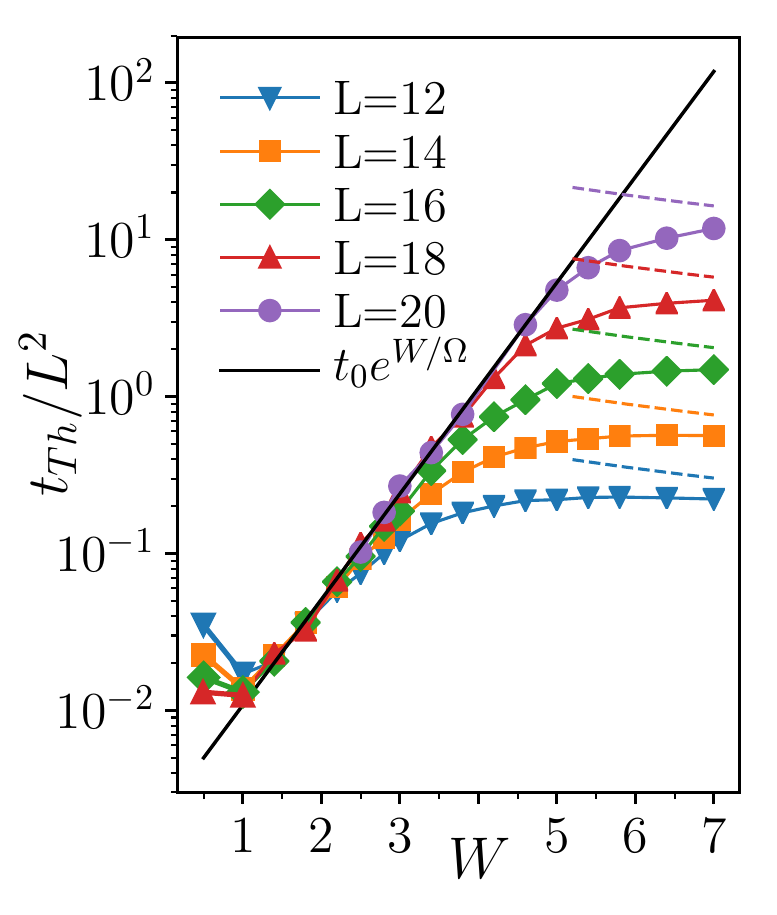}\hspace{-0.2cm}\includegraphics[width=0.5\linewidth]{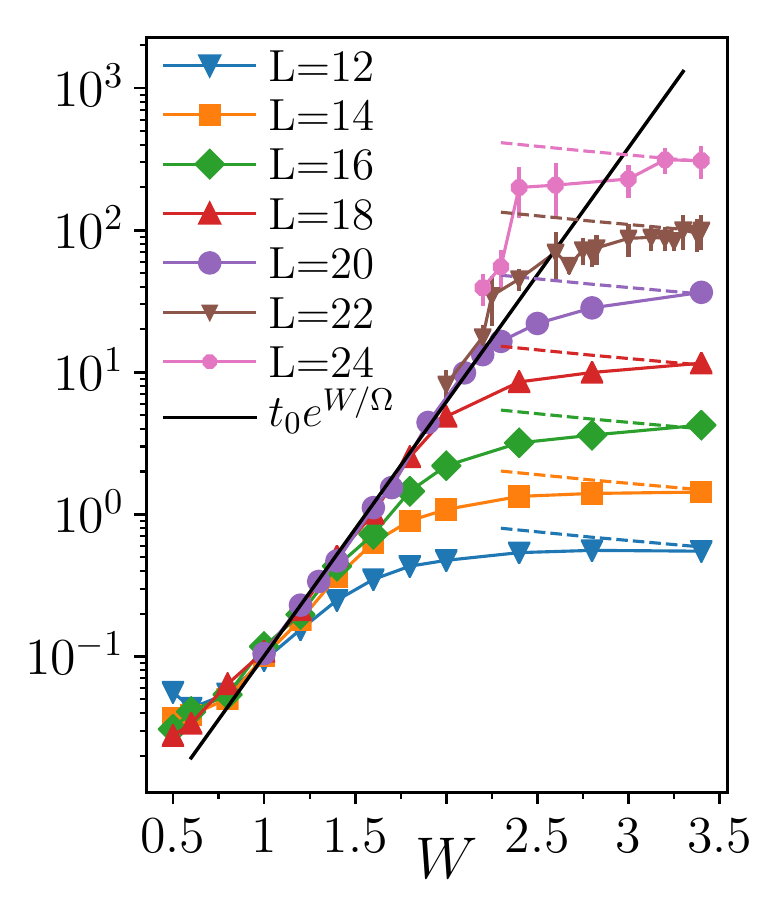}
 \vspace{-0.6cm}
  \caption{ { Thouless time $t_{Th}$ for $J_1-J_2$ model (left) and XXZ model
  (right)  extracted from the SFF.
 $t_{Th}$ is divided by $L^2$ to
  emphasize the scaling with system size $L$.
  Dashed lines show the Heisenberg time $t_H$.
 }} \label{j1j2SFF} 
\end{figure}

  In the case of $J_1-J_2$ model, Thouless times obtained for available system sizes 
 seem to follow the scaling \eqref{ThSc} as for increasing
 system size $L$, the point $\tilde W(L)$ where $t_{Th}/L^2(W)$ deviates
 from the 
 $e^{W/\Omega}$ behavior shifts to larger disorder strength, as shown in the left panel of Fig.~\ref{j1j2SFF}. 
 An interpretation of this behavior along the lines of \cite{Suntais19}
 is that one assumes that system size dependence of
 $\tilde W(L)$ continues indefinitely, so that the scaling \eqref{ThSc} holds in the 
 thermodynamic limit. This would imply that there is no transition to a MBL phase.
 However, the Thouless time scaling obtained for available system sizes in the 5D Anderson model,
 exhibited in 
 {the lower panel of} Fig.~\ref{andtTh}, is very similar with larger system sizes deviating from 
 \eqref{ThSc} at larger disorder strength. As such a behavior occurs in the 5D Anderson model
 despite the localization transition taking place at $W^{5D}_C$,
 we may give a second possible interpretation of the result: the scaling \eqref{ThSc} is 
 not broken at available system sizes
 because of
 strong finite size effects. While it is still possible to devise the location of the critical point $W^{5D}_C$
 provided one knows the correct value of the exponent $\alpha$ governing the subdiffusion at 
 the Anderson transition, it is not clear how to rescale the
 Thouless times $t_{Th}$ in the many-body case since the transport properties on the delocalized side
 are not fully understood, with 
 a suggestion of subdiffusive behavior 
 with exponent $\alpha$ vanishing close to the transition \cite{Luitz17b}. Presumably, a sensible 
 criterion for the transition in the many-body case would be $t_{Th}\propto t_H\propto e^{cL}$.
In any case, the main observation in \cite{Suntais19} is that $t_{Th}/L^2$ is approximately equal to $e^{W/\Omega}$ in the deeply 
{delocalized} regime {of the $J_1-J_2$ model}.
This implies, in turn, that the diffusion coefficient $D(W)$ decreases like  $e^{-W/\Omega}$, exactly like in the 3D and 5D Anderson models.
Concluding that  {$D(W)$} never vanishes is a dangerous extrapolation, which leads to incorrect results for the Anderson models.
{The similarity of Thouless time scaling for 5D Anderson and $J_1-J_2$ models suggests that 
the conclusion of \cite{Suntais19} about $D(W)\propto e^{-W/\Omega}$ in the $J_1-J_2$ model for any disorder strength in the thermodynamic limit is misleading. Our results show that the apparent 
scaling \eqref{ThSc} is probably a finite size effect.

 The finite size effects in the $J_1-J_2$ model are necessarily enhanced by the next-to-nearest neighbor coupling term, thus we may expect weaker finite size effects for the XXZ model. The scaling of Thouless time for {this model}  is presented in the right panel of Fig.~\ref{j1j2SFF} and it follows \eqref{ThSc} only for disorder strengths $W\in[1,2]$.
We observe two important differences with the results for $J_1-J_2$ model. Firstly, at weak disorder $W$, 
 the exponential dependence of the Thouless time $t_{Th}$ on $W$ is weaker than in the interval $W\in[1,2]$. This is due to the proximity of the integrable point $W=0$ \cite{Bethe31,Alcaraz87} with Poisson level statistics and $t_{Th}=t_H$. 
 Secondly and more importantly, we see a breakdown of \eqref{ThSc} for the XXZ model at $W\gtrsim2$ where the data for 
 $L=22$ and $L=24$ exceed the $t_0 e^{W/\Omega}$ line even though the Thouless time is still an order 
 of magnitude smaller than the Heisenberg time $t_H$. This indicates that the exponential scaling with $W$ is a numerical
observation {explicitly broken in the XXZ model and likely valid only} in a limited range {in other systems.}  The data for $L=22$ and $L=24$ are available only for 
 $W\geq2$ and $W>2.2$ \cite{suppl}. Nevertheless, the breakdown of the scaling \eqref{ThSc} for $L=22,24$ at $W\approx 2.2$
 is apparent, {indicating} that the $L^2$ scaling of Thouless time breaks down. This reflects the slow-down of transport 
 and approaching the MBL transition when $t_{Th}\propto t_H\propto e^{cL}$.

 {\it Conclusions. } 
 Our results show that the Thouless time, defined by the behavior of the SFF,
 reflects the transport properties in disordered non-interacting models as we have shown on the examples of 3D and 5D Anderson models. In particular, the scaling of the Thouless time $t_{Th}$ at the transition encodes the subdiffusive behavior of the mean square displacement $\langle r^2(t) \rangle \sim t^{\alpha}$ with the exponent $\alpha_{3D}=2/3$ and $\alpha_{5D}=2/5$ leading to scaling $t_{Th} \sim L^{2/\alpha}$ with system size at the transition.

  The scaling of Thouless time for $J_1-J_2$ model seems to follow $t_{Th} \sim t_0 L^2 e^{W/\Omega}$, however, the behavior of $t_{Th}$ is directly analogous to the case of 5D Anderson model.
  The latter undergoes a transition to a localized phase and the Thouless time does not exceed the $t_0 L^2 e^{W/\Omega}$ curve only because of strong finite size effects at available system sizes. 
  It is plausible that the situation is the same in 
  the $J_1-J_2$ model, raising doubts about the claims of \cite{Suntais19}.
  Our results for XXZ model demonstrate that the $L^2$ scaling of the Thouless time $t_{Th}$, valid deep in delocalized phase is evidently broken at $W\approx 2.2$, signaling a transition to a MBL phase at a strong disorder.

Finally, let us mention alternative definitions of Thouless time
 \cite{Beugeling15,Serbyn17, Torres-Herrera15, TorresHerrera17, Torres-Herrera18, Schiulaz19, Schiulaz19b}.
 Comparison of these different approaches is in progress.
   While finalizing this manuscript, we became aware of the
  related works \cite{Abanin19a, Panda20}.

  {\it Acknowledgments. } 
  We are most grateful to Fabien Alet for kindly sharing with us the eigenvalues for $L=22,24$ XXZ model as well as discussions
  on subjects related to this work.
  The computations have been performed within PL-Grid Infrastructure, its support is acknowledged. We acknowledge the support of 
  National Science Centre (Poland) under projects  2015/19/B/ST2/01028 (P.S. and J.Z.), 2018/28/T/ST2/00401 
  (doctoral scholarship -- P.S.) as well as Polish-French bilateral grant Polonium 40490ZE.


%

\newcommand{\snum}{S}

\renewcommand{\theequation}{\snum.\arabic{equation}}

\setcounter{equation}{0}

 \pagebreak

 \section{Supplementary: extracting Thouless time from Spectral Form Factor}
 \label{appendix1}

 The spectral form factor (SFF) is the tool employed in analysis of level statistics in Anderson models and of disordered quantum spin chains presented in the main text. In this supplementary material we recall definition of SFF, provide details of performed numerical calculations and describe the employed method of extraction of the Thouless time from SFF.
 
  The spectral form factor (SFF) is defined as
      \begin{equation}
 K(\tau)  = \frac{1}{Z}\left \langle \left| \sum_{j=1}^{\mathcal N} g( \epsilon_j) \mathrm{e}^{-i \epsilon_j \tau} \right|^2 \right \rangle,
 \vspace{-0.1cm}
 \label{eq: KtS}
\end{equation}
where $\mathcal N$ is the dimension of Hilbert space.
The eigenvalues  $\epsilon_j$ 
are obtained in the so called unfolding procedure. During the unfolding, a 
level staircase 
function $\sigma(E)= \sum_i \Theta(E- E_i)$ (obtained from the set of eigenvalues of the system $\{E_i\}$ ordered in an ascending manner)
is separated into smooth and fluctuating parts $\sigma(E) = 
\overline \sigma(E) + \delta \sigma (E)$ and the eigenvalues are mapped via 
\begin{equation}
 E_j \rightarrow \epsilon_j
= \overline \sigma(E_j).
\label{eq: unf1}
\end{equation}
As the smooth part $\overline \sigma(E)$ we take a polynomial of certain
small degree $n_p$ fitted to the level staircase function $\sigma (E)$.
To calculate SFF we use
$g(\epsilon) \propto \exp( -(\epsilon -\bar \epsilon)^2/{2 \eta \sigma_{\epsilon}}^2 )$, where $\bar \epsilon$ denotes the average of the unfolded eigenvalues for given disorder realization $\epsilon_i$, $\sigma_{\epsilon}$ is the standard deviation of $\{ \epsilon_i\}$ and $\eta = 0.3$. This choice of parameters follows precisely \cite{Suntais19}.

The summation in \eqref{eq: KtS} extends over the whole spectrum of the system. 
To get all of the eigenvalues we perform exact diagonalization (ED) for Anderson models in 3D and 5D. We calculate SFF according to formula \eqref{eq: KtS}, using unfolding with polynomial of order $n_p=10$
and averaging results over more than 400 disorder realizations. 
Similarly, we perform ED of disordered Heisenberg spin chain and $J_1-J_2$ model of size
$L \leqslant 16$ ($L=18$) and average results over $1000$ ($500$) disorder realizations.

An exemplary result for SFF of 3D Anderson model is shown in Fig.~\ref{3dAndKt}. One clearly observes a value $\tau_{Th}=t_{Th}/t_{H}$ beyond which SFF of 3D Anderson model follows the GOE prediction. 
To quantitatively extract Thouless time $t_{Th}$ from the SFF, we follow \cite{Suntais19} and consider a function
 \begin{equation}
  \Delta K(t/t_H) = \left| \log\left( \frac{K(t/t_H)}{K_{GOE}(\tau=t/t_H)}\right)\right|.
 \end{equation}
 The Thouless time $t_{Th}$ is the smallest positive time for which $\Delta K(t/t_H)< \varepsilon$. We choose the value of cut-off $\epsilon = 0.05$. The choice of cut-off $\epsilon$ affects the obtained values of Thouless time $t_{Th}$, however, reasonable changes in value of $\epsilon$ do not affect the obtained scaling of Thouless time with system size and disorder strength.
 \begin{figure}
 \includegraphics[width=0.92\linewidth]{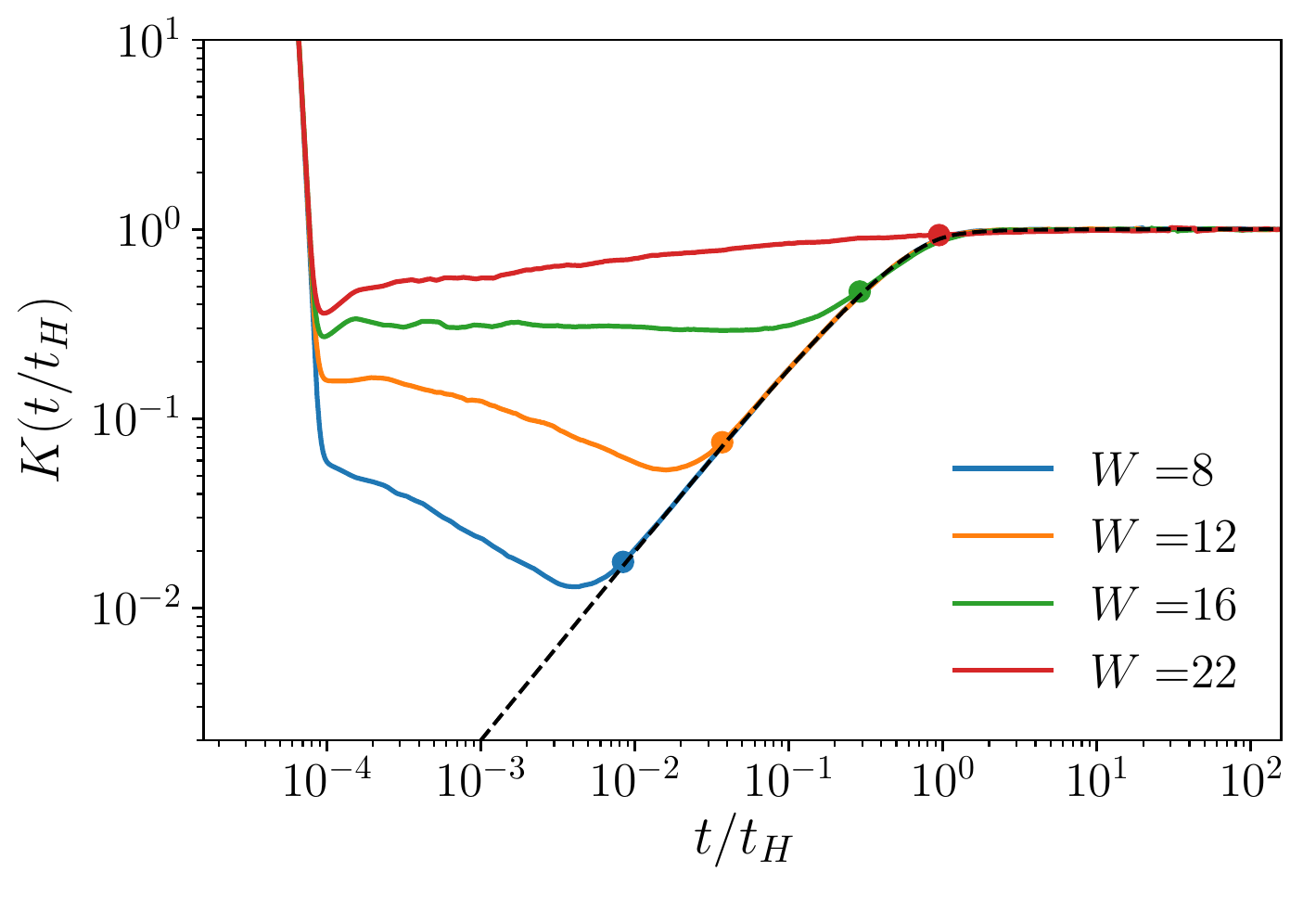}
  \caption{ {Spectral form factor $K(\tau)$ for 3D Anderson model of size $L=40$, dots denote obtained values of Thouless time. The black dashed lines denote the SFF of GOE ensemble which is known analytically.
 }} \label{3dAndKt}
\end{figure} 
 
 The eigenvalues for larger sizes  $L=20, 22, 24$ of considered spin chains
 can be obtained with the shift-and-invert method. However, the shift-and-invert method
 provides only a certain number $n_e$ eigenvalues around a specific target
 energy. Using $n_e=2000, 100, 50$ eigenvalues from the middle of system spectrum for system size $L=20, 22, 24$ and averaging results over more than $200$ disordered realizations we have verified that the SFF can still be calculated if the available $n_e$ eigenvalues is taken into account in the 
 sum \eqref{eq: KtS}. In those cases we have used unfolding with polynomial of degree $n_p=3$ to avoid over-fitting of the staircase function \cite{Gomez02}. If a small fraction $n_e/\mathcal N$ of eigenvalues is considered in the sum \eqref{eq: KtS}, the SFF is correctly reproduced only for $t>t_M$ with $t_M$ depending on $n_e$ as well as on the system size $L$. This can be intuitively understood: since $t_{Th}$ is inversely proportional to $E_{Th}$ -- the energy scale at which the correlations of eigenvalues are well described by GOE, the upper bound on energy scales probed by SFF is proportional to the number of available eigenvalues $n_e$ which translates to the lower bound $t_M$ on accessible times. This shows that the Thouless times can be extracted from the fraction of eigenvalues from shift-and-invert method if the disorder is sufficiently strong.

\end{document}